\documentclass{article}
\usepackage{amsfonts,amssymb, amsmath}
\usepackage[english]{babel}
\usepackage{hyperref}

\newtheorem{ex}{Example}

\begin{document}


\title{  Reduction of divisors and  Clebsch system}

\author{ A.V. Tsiganov\\
Steklov Mathematical Institute of Russian Academy of Sciences,\\
 Moscow, Russia\\
andrey.tsiganov@gmail.com
}

\date{}
\maketitle

\begin{abstract}
There are a few Lax matrices of the Clebsch system. Poles of the Baker-Akhiezer function determine the class of equivalent divisors on the corresponding spectral curves. According to the Riemann-Roch theorem, each class has a unique reduced representative. We discuss properties of such reduced divisor on the spectral curve of $3\times 3$ Lax matrix having a natural generalization to  $gl^*(n)$ case.
\end{abstract}

\section{Introduction}

The Clebsch's system was proposed in 1870 and it represents a specific famous case of the Kirchoff equations which describes the motion of a rigid body in an ideal fluid  \cite{cl70}.
A full description of the long history of  this system can be found in the book of Borisov and Mamaev pages 218-220, \cite{bm}. We will limit ourselves to listing a few points from this book.

The Clebsch's system  is isomorphic or belongs to a few families of integrable systems:
\begin{itemize}
 \item  rigid body motion in a central Newtonian field, see  textbooks \cite{ar77,bm} and papers by Brun \cite{brun93},  Tisserand \cite{tiss72} and Weber \cite{web78} about integrable electrodynamic systems;
\item geodesic motion on ellipsoid \cite{min}  and Kowalevski gyrostat  \cite{kt05};
 \item  Frahm- Schottky-Manakov system on $so(4)=so(2)\times so(2)$, see \cite{bob86,ts06a};
 \item  Landau-Lifshits equation and antisymmetric chiral $O(3)$ field model, see \cite{cher81,ves83} and references in \cite{bob87};
 \item  integrable systems on  $gl^*(n, R)$ obtained by Perelomov, see \cite{p81};
\item  $n$-cite elliptic Gaudin model  on $so(2)\times so(2)\cdots\times so(2)$, see \cite{skl98} and references in \cite{bob87};
\item quasi-St\"{a}ckel integrable systems \cite{sok05,sok08}.
\end{itemize}
So, we have a set of equivalent two-dimensional integrable systems appearing in the different physical applications.  Let us briefly describe known
approaches to solution of the corresponding equations of motion:
\begin{itemize}
\item  in 1859 Neumann integrated equations of motion  of a point on the sphere which coincide with Clebsch's equations of motion in a partial case  \cite{neum59}
\item in 1879 Weber solved Clebsch's equations of motion for the Neumann partial case  \cite{web78};
\item in 1887 Halphan   solved Clebsch's equations in terms of elliptic functions appearing in  another partial case  \cite{hal};
  \item in 1891  Schottky integrated  the Frahm- Schottky-Manakov equations of motion  \cite{shot91};
  \item in 1892 K\"{o}tter solved Clebsch's equations of motion in generic case   \cite{kot92};
  \item in 1895 Kobb obtained  quadratures  for  a rigid body motion in a central Newtonian field \cite{kobb}. Construction of these quadratures  involves
  solution of the fourth order algebraic equation;
   \item in 1893 Steklov studied  quadratures in generic and particular cases of Clebsch's system \cite{stek93,stek02};
  \item in  1900 Chaplygin studied  particular solutions of  Clebsch's equations using characteristic function theory \cite{chap90};
   \item In 1959 Harlamova obtained  quadratures generalising Chaplygin's method in the generic case. These quadratures also involve solutions of the fourth order algebraic equation \cite{har59};
  \item in 1963 and 1974 Arkhangel'skii, Demin and Kiselev studied periodic solutions of Clebsch's equations  under certain restrictions \cite{ar77,dem74};
  \item in 1987 Bobenko presented theta-functions formulae for all the classical tops using finite-gap theory and $2\times 2$ Lax matrices for the Clebsch and  Frahm- Schottky-Manakov systems \cite{bob87};
  \item in  1998 Zhivkov and Christov presented  theta-functions formulae using finite-gap theory and  $4\times 4$ Lax matrices  \cite{ziv98};
  \item in 1998 Sklyanin and Takebe obtained variables of separation for the elliptic Gaudin magnet \cite{skl98};
\item in  2008 Sokolov and Marikhin obtained quadratures for a pair of quasi-St\"{a}ckel Hamiltonians \cite{sok05, sok08}
\[H_1=ap_1^2+2b p_1p_2+cp_2^2+dp_1+ep_2+f\]
\[H_2=Ap_1^2+2Bp_1p_2+Cp_2^2+Dp_1+Ep_2+F\,.\]
 In this case the construction of quadratures involves solutions of the cubic algebraic equation;
\item in 2015 Magri and Skrypnyk found quadratures for the complex  variables of separation, which are two roots of the cubic algebraic equation \cite{mag15};
\item in 2021 Fedorov, Magri and Skrypnyk solved Clebsch system in theta-function using complex variables of separation, which consist of one of the eight solutions of the system of quadratic algebraic equations and its derivative \cite{mag21,mag21a}.
\end{itemize}
Following to Borisov and Mamaev \cite{bm} we  fully agree with Chaplygin \cite{chap90} and  Magnus \cite{mag} that these solutions of equations of motion belong rather to the field of mathematical sport and add nothing to the description of motion. Nevertheless, it is natural to assume that this list will be continued and, one day, we will see unambiguously defined real  variables of separation depending on real time, similar to the Kowalevski top \cite{kow89} and Euler's two fixed centers problem \cite{bis16}.

In this note we do not study  solutions of equations of motion in theta-functions obtained via quadratures or  without quadratures. We also do not  discuss relations between different quadratures. Our aim is to study properties of the Lax matrix  for integrable systems on  $gl^*(n, R)$ obtained by Perelomov \cite{p81}. For all these matrices, number degrees of freedom $n$ is more then genus of the corresponding spectral curve, similar to the Heisenberg  and  Gaudin magnets.

For instance, when $n=3$ the corresponding Baker-Akhiezer vector function $\psi$ has three poles $P_1,P_2$ and $P_3$ on the genus one spectral curve.  So, there is a chain of equivalent divisors
\[D= P_1+P_2+P_3\to D'=P'_1+P'_2\to \rho(D)=D''=P''\]
according to the  Riemann-Roch theorem for divisors on algebraic curves \cite{mir}.  Below we calculate all these points on a spectral curve using standard Abel's reduction algorithm \cite{ab} and prove that:
\begin{itemize}
  \item affine coordinates of points $P'_{1,2}$ are functions on integrals of motion and one variable $X$, which also can be found in the Kobb and Harlamova quadratures;
  \item affine coordinates of the point $P''$ in the reduced divisor $\rho(D)$ are functions only on integrals of motion.
\end{itemize}
Thus, we prove that reduced divisor for  the $gl^*(n, R)$ system at $n=3$ has the same properties as reduced divisors  for  the harmonic oscillator and Kepler problem \cite{ts20rr}, for the Drach systems \cite{ts20ds}, for the Heisenberg  and  Gaudin magnets \cite{ts20m}, for the   Kowalevski top \cite{ts20} and the Euler two fixed centers problem  \cite{ts21}.

The main result is to prove for $3\times 3$ Lax matrix of  the Clebsch system the non-obvious assumption that reduced divisor is fixed, i.e. independent on time, when the dimension of the configurational space is more then genus of the corresponding spectral curve.

\subsection{ Kirchhoff equations in the Clebsch case}
 In the Clebsch case,  the Kirchhoff equations for the motion of a rigid body in an  ideal incompressible fluid are equal to
\begin{equation}\label{eq-clm}
\dot {M}=p\times A\,p\,, \qquad
\dot {p}= p \times M\,.
\end{equation}
Here $p$ and $M$ are three-dimensional vectors, $\times$  denotes the vector product and $A$ is a diagonal matrix
\[
A=\left(
    \begin{array}{ccc}
      a_1 &0 & 0 \\
      0 & a_2 & 0 \\
      0 & 0 & a_3 \\
    \end{array}
  \right)\,,\qquad a_i\in\mathbb R\,.
\]
Vector $M$ is the total angular momentum vector, whereas $p$ represents the total linear momentum of a system \cite{cl70}.  There are two geometric integrals of motion
\[
c_1=p_1^2+p_2^2+p_3^2\,,\qquad c_2=p_1M_1+p_2M_2+p_3M_3
\]
and two Hamiltonians
\begin{equation}\label{ham}\begin{array}{rcl}
H&=&  M_1^2 + M_2^2 + M_3^2+a_1(p_2^2 + p_3^2) + a_2(p_1^2 +p_3^2) + a_3(p_1^2 + p_2^2)\,,
\\ \\
K&=& a_1M_1^2 + a_2M_2^2 +a_3 M_3^2+ a_2a_3p_1^2 + a_1a_3p_2^2 +a_1a_2p_3^2\,,
\end{array}
\end{equation}
which Poisson commutes with respect to Lie-Poisson bracket on the Lie algebra $e^*(3)$
\begin{equation}\label{e3}
\,\qquad \bigl\{M_i\,,M_j\,\bigr\}=\varepsilon_{ijk}M_k\,, \qquad
\bigl\{M_i\,,p_j\,\bigr\}=\varepsilon_{ijk}p_k \,, \qquad
\bigl\{p_i\,,p_j\,\bigr\}=0\,,
\end{equation}
where $\varepsilon_{ijk}$ is a skew-symmetric tensor. All these definitions are invariant under the cyclic permutation of indexes.

We know reductions of these equations of motion (\ref{eq-clm}) to quadratures,  solutions of these equations in terms of theta-functions,   bi-Hamiltonian structure of these equations   \cite{ts06,ts07}, various  Lax pair representations  \cite{avm88, bbe94,p81,ziv98},   topological invariants \cite{bf}, Hirota-Kimura type discretization \cite{pet11}, numerical solutions \cite{pop19}, etc.

\section{Lax matrix and spectral curve}
\setcounter{equation}{0}
If we know Lax representation of the original  equations of motion (\ref{eq-clm})
 \begin{equation}\label{lax-g}
\frac{d}{dt}\mathcal  L(x)=[\mathcal  L(x), \mathcal  A(x)]\,,
\end{equation}
for two $N\times N$ matrix functions  $\mathcal  L(x)$ and $\mathcal  A(x)$ on  phase space depending on  the auxiliary spectral parameter $x$, then  we can directly integrate equations (\ref{lax-g}) in terms of theta-functions by using finite-gap integration theory. Indeed, the time-independent spectral equation
\begin{equation}\label{ba-f}
\mathcal  L(x)\, \psi(x,y) = y\, \psi(x,y)
\end{equation}
allows us to represent the  Baker-Akhiezer  vector function $\psi$ in terms of the Riemann theta function on a nonsingular compactification of the spectral curve defined by  equation
\[
\Gamma:\qquad f(x,y)=\det(L(x)- y)=0\,.
\]
The second equation defines time
\begin{equation}\label{time-eq}
\frac{d}{dt}\psi(x,y)=-\mathcal  A(x)\psi(x,y)\,,
\end{equation}
 and evolution of original variables  $p_i(t)$ and $M_i(t)$ with respect to this time, see details in \cite{bob86,ziv98}.

For integrable systems on   $gl^*(n, R)$ Lax representation (\ref{lax-g}) 	 was found by Perelomov  \cite{p81}. When $n=3$ we have the following   $3\times 3$
\begin{equation}\label{lax}
 \mathcal  L(\lambda)=A+\lambda \mathrm M+\lambda^2 \,p\otimes p=
 \left(
   \begin{array}{ccc}
     \lambda^2p_1^2+a_1 & \lambda^2p_1p_2 + M_3\lambda  & \lambda^2p_1p_3 - M_2\lambda  \\
     \lambda^2p_1p_2 - M_3\lambda  & \lambda^2p_2^2+a_2 & \lambda^2p_2p_3 + M_1\lambda  \\
     \lambda^2p_1p_3 + M_2\lambda & \lambda^2 p_2p_3 - M_1\lambda  & \lambda^2p_3^2 +a_3
   \end{array}
 \right)\,,
\end{equation}
Here $\mathrm M\in so^*(3)$ is a skew-symmetric matrix associated with  vector $M\in \mathbb R^3$ and one of the corresponding second Lax matrices
\[
\mathcal A=-\frac{2}{\lambda}\left(A+\lambda \mathrm M\right)\qquad\mbox{or}\qquad \mathcal A=2\lambda  \,p\otimes p\,.
\]
in (\ref{lax-g}) defines equations of motion coinciding with  the Kirchhoff equations (\ref{eq-clm}).

\subsection{Poles of the  Baker-Akhiezer function}
Let us  emphasize again that we do not discuss  equations of motion (\ref{eq-clm}) or (\ref{lax-g}) et al, because instead of evolution of poles of the  Baker-Akhiezer function we want to study reduction of poles of the  Baker-Akhiezer function only.

Indeed, spectral curve  $\Gamma$
\begin{equation}\label{s-curve}
\Gamma:\qquad f(\mu,\lambda)= c_2^2\lambda^4 + (c_1\mu^2 - H\mu + K)\lambda^2 + \det(A-\mu)=0
\end{equation}
 is a 2-fold covering of elliptic curve $E$ at $\lambda^2=y$
\begin{equation}\label{e-curve}
E:\qquad f(\mu,y)= c_2^2y^2 + (c_1\mu^2 - H\mu + K)y + \det(A-\mu)=0\,.
\end{equation}
Reduced divisor   is a point $P''$ on  this elliptic curve $E$ (\ref{e-curve}).  Our aim is calculate this point and study its affine coordinates.

As a first step in this direction we have to  construct a class of linearly  equivalent divisors
\[D=\sum_{i=1}^m P_i\,,\]
which are a formal sum of poles $P_i=(\mu_i,\lambda_i)$  of the  vector Baker-Akhiezer function $\psi $ (\ref{ba-f})  with some fixed normalization $\vec\alpha$
\[
\vec{\alpha} \cdot \psi=\sum_{i=1}^N \alpha_i\psi_i=1\,,\
\]
Because
\[
\psi_j=\dfrac{(L(\lambda)-\mu)^\wedge_{jk}}{(\vec\alpha\cdot (L(\lambda)-\mu)^\wedge)_k}\,,\qquad
\forall k=1,2,3\,,
\]
where the wedge denotes the adjoint or co-factor matrix,  poles of the Baker-Akhiezer function $\psi(\lambda,\mu)$ are  common zeroes of the four  algebraic equations
\begin{equation}\label{m-eq}
f(\mu,\lambda)=\det(L(\lambda)- \mu)=0\qquad\mbox{and}\qquad \vec\alpha\cdot (L(\lambda)-\mu)^\wedge)=0\,.
\end{equation}
For  general normalization $\vec\alpha$ algebraic equations (\ref{m-eq}) have five solutions $P_i=(\mu_i,\lambda_i)$, $i=1..5$, whereas  at \[\vec{\alpha}=(p_1,p_2,p_3)\]
there are only three solutions of equations (\ref{m-eq}). Indeed, three solutions of equations (\ref{m-eq}) define three points $P_i=(\mu_i,\lambda_i)$ on  spectral curve $\Gamma$ (\ref{s-curve}) which form positive or semi-reduced divisor
\[D=P_1+P_2+P_3\,, \quad \mbox{deg}D=3\]
on the genus three  algebraic curve $\Gamma$, $g(\Gamma)=3$. According to  the Riemann-Roch theorem, the dimension of
 linear system $|D|$, which  is the set of all the nonnegative divisors  linearly equivalent to $D$
 \[
|D|=\{D'\in \mathrm{Div}(\Gamma)\,|\,D'\sim D\,\mathrm{and}\,D'>0\}\,,
\]
is equal to
\[\mathrm{dim} |D|=\mbox{deg}\,D-\mbox{genus}(\Gamma)=3-3=0\,,\]
see definitions and other details in textbook \cite{mir}.

In our case, dim$|D|=0$ and, therefore,   divisor $D$ is a unique reduced divisor in the corresponding class of equivalent divisors. Thus, we can not reduce this divisor to prime divisor of degree one on a genus three bielliptic curve $\Gamma$.

On  elliptic curve $E$ three points $P_i=(\mu_i,y_i)$, where $\lambda_i^2=y_i$,  define  semi-reduced  divisor on the genus one elliptic curve $E$ (\ref{e-curve})
\[D=P_1+P_2+P_3\,, \quad \mbox{deg}D=3\]
so that
\[
\mbox{dim}|D|=\mbox{deg}D-\mbox{genus}(E)=3-1=2\,.
\]
According to the Riemann-Roch theorem, we can reduce this semi-reduced divisor $D$, to equivalent divisors $D'$ and $D''$:
\[
D\to D'\to D''\,,\qquad \mbox{dim}|D|=2\,,\quad \mbox{dim} |D'|=1\,,\qquad \mbox{dim}|D''|=0\,.
\]
Below we study  these semi-reduced $D'$ and reduced $D''=\rho(D)$ divisors on the elliptic curve $E$ (\ref{e-curve}).

\subsection{Semi-reduced divisor of degree three}
For generic normalization $\vec{\alpha}$ the last  three equations in (\ref{m-eq}) are cubic polynomials in $\lambda$
\[
e_i=(\vec\alpha\times  p)_i\,\lambda^3+e_i^{(2)}(\mu)\lambda^2+e_i^{(1)}(\mu)\lambda+e_i^{(0)}(\mu)=0\,,\quad i=1,2,3
\]
and solutions of  these equations for $\lambda$ and $\mu$ are roots of fifths order polynomials
\[
A_5(\lambda)=0\,,\qquad B_5(\mu)=0\,,
\]
which can be easily obtained using modern computer algebra systems and, therefore, here we do not present these polynomials for brevity.

If $\vec{\alpha}=(p_1,p_2,p_3)$, then vector $(\vec\alpha\times  p)=0$ is equal to zero and algebraic equations (\ref{m-eq}) have the following form
\[
\begin{array}{l}
f(\mu,\lambda)=c_2^2\lambda^4 + (c_1\mu^2 - H\mu + K)\lambda^2 + \det(A-\mu)=0\,,\\
\\
e_1=M_1c_2\lambda^2 +\bigl((M_2p_3 - M_3p_2)\mu+ a_3M_3p_2 - a_2M_2p_3 \bigr)\lambda + p_1\bigl(\mu^2 -(a_2 + a_3)\mu + a_2a_3\bigr)=0,\\
\\
e_2=M_2c_2\lambda^2 + \bigl((M_3p_1-M_1p_3)\mu + a_1M_1p_3 - a_3M_3p_1 \bigr)\lambda + p_2\bigl(\mu^2 - (a_1 + a_3)\mu + a_1a_3\bigr)=0,\\
\\
e_3=M_3c_2\lambda^2 + \bigl((M_1p_2 - M_2p_1)\mu + a_2M_2p_1-a_1M_1p_2 \bigr)\lambda + p_3\bigl(\mu^2 - (a_1 + a_2)\mu + a_1a_2\bigr)=0.
\end{array}
\]
Solutions of these equations for $\lambda$ and  $\mu$ are   roots of the cubic polynomials
\begin{equation}\label{a-gen}
\begin{array}{rcl}
A_3(\lambda)&=&-b_3c_2\lambda^3
+\Bigl(a_1(p_2M_3-p_3M_2)(c_1M_1 + c_2p_1) \Bigr.\\
\\&+&\Bigl. a_2(p_3M_1 -p_1 M_3)(c_1M_2 + c_2p_2) + a_3(p_1M_2-p_2M_1)(c_1M_3 + c_2p_3)\Bigr)\lambda^2\\
\\
&-&\Bigl((p_2^2 + p_3^2)(a_1 - a_3)(a_1 - a_2)p_1M_1 + (p_1^2 + p_3^2)(a_2 - a_3)(a_2 - a_1)p_2M_2\Bigr.\\
\\
 &+&\Bigl. (p_1^2 + p_2^2)(a_3 - a_2)(a_3 - a_1)p_3M_3\Bigr)\lambda
-p_1p_2p_3(a_1-a_2)(a_2 - a_3)(a_3 - a_1)
\end{array}
\end{equation}
and
\begin{equation}\label{b-gen}
B_3(\mu)=b_3\mu^3+b_2\mu^2+b_1\mu+b_0\,,
\end{equation}
where
\[\begin{array}{rcl}
b_3&=&c_1(M_1^2+M_2^2+M_3^2)-c_2^2\,,\\
\\
b_2&=&2c_2(a_1p_1M_1+a_2p_2M_2+a_3p_3M_3)-2c_1(a_1M_1^2+a_2M_2^2+a_3M_3^2)\\
\\
&-&a_1 (M_2p_3 - M_3p_2)^2 - a_2 (M_1p_3 - M_3p_1)^2 -a_3 (M_2p_1 - M_1p_2)^2\,,\\
\\
b_1&=&(a_1a_2 + a_1a_3 + a_2a_3)b_3 - p_1^2(a_2 + a_3)\Bigl((a_1 - a_2)M_2^2 + (a_1 - a_3)M_3^2\Bigr)\\
 \\
 &-& p_2^2(a_1 + a_3)\Bigl((a_2 - a_1)M_1^2 + (a_2 - a_3)M_3^2\Bigr)
      - p_3^2(a_1 + a_2)\Bigr((a_3 - a_1)M_1^2 + (a_3-a_2)M_2^2\Bigr)\,,\\
\\
b_0&=&a_1a_2a_3c_2^2 - (a_1M_1^2 +a_2 M_2^2 +a_3 M_3^2)( a_2a_3p_1^2+ a_1a_3p_2^2 +a_1a_2p_3^2 )\,.
\end{array}
\]
The roots of polynomials $A_3(\lambda)$ and $B_3(\mu)$  determine poles $P_i=(\mu_i,\lambda_i)$ of the Baker-Akhiezer function on
 spectral bielliptic curve $\Gamma$ (\ref{s-curve}).

To determine the corresponding poles $P_i=(\mu_i,y_i=\lambda_i^2)$ on elliptic curve $E$ we can replace three equations depending on $\mu$ and $\lambda$
\[e_k(\mu,\lambda)=e_k^{(2)}\lambda^2 +e_k^{(1)}\lambda+e_k^{(0)}=0\,,\qquad k=1,2,3\,,\]
with three equations depending  on $\mu$ and $y=\lambda^2$
\[\begin{array}{c}
E_{12}(\mu,y)=e_2^{(1)}e_1-e_1^{(1)}e_2=0\,,\\
E_{23}(\mu,y)=e_3^{(1)}e_2-e_2^{(1)}e_3=0\,,\\
E_{31}(\mu,y)=e_1^{(1)}e_3-e_3^{(1)}e_1=0\,.
\end{array}
\]
Equations of motion for the Clebsch system (\ref{eq-clm}), Hamiltonians $H_{1,2}$ (\ref{ham}) and polynomial $B_3(\mu)$ are
invariant under cyclic permutations of the indexes. Thus, we consider a linear combination of  $E_{ik}$
 \[
\mathrm g(\mu,y)=\sum\varepsilon_{ijk} p_iE_{jk}=p_1E_{23}+p_2E_{31}+p_3E_{12}=c_2\mathcal Q(\mu)y-\mathcal P(\mu)=0\,,
\]
which is also invariant under permutations. Here $\mathcal Q(\mu)$ and $\mathcal P(\mu)$ are polynomials of first and second order in $\mu$
\[
\mathcal Q(\mu)=-\mu b_3+(a_1M_1^2 + a_2M_2^2 +a_3 M_3^2)c_1 - c_2X\,,\qquad
\mathcal P(\mu)=\mathcal P_2\mu^2+\mathcal P_1\mu+\mathcal P_0\,,
\]
where
\begin{equation}\label{x-coord}
X=a_1p_1M_1 + a_2p_2M_2 + a_3p_3M_3\,.
\end{equation}
and
\[\begin{array}{rcl}
\mathcal P_2&=&(a_1p_1^2 + a_2p_2^2 + a_3p_3^2)c_2 - c_1X\,,\\
\mathcal P_1&=&\Bigl((a_1 + a_2 + a_3)c_1 - a_1p_1^2 - a_2p_2^2 - a_3p_3^2\Bigr)X\\
&-& \Bigl((a_1a_2 + a_1a_3 + a_2a_3)c_1 - (a_1a_2p_3^2 + a_1a_3p_2^2+a_2a_3p_1^2)\Bigr)c_2\,,\\
\mathcal P_0&=&a_1a_2a_3c_1c_2 - (a_1a_2p_3^2 + a_1a_3p_2^2 + a_2a_3p_1^2)X\,.
\end{array}
\]
It is easy to prove that three points $P_i=(\mu_i,y_i)$ with coordinates
\[
B_3(\mu)=b_3(\mu-\mu_1)(\mu-\mu_2)(\mu-\mu_3)=0\quad\mbox{and}\quad
y_i=\frac{\mathcal P(\mu_i)}{c_2\mathcal Q(\mu_i)}\,,\qquad i=1,2,3,
\]
lie on  elliptic curve $E$ (\ref{e-curve}). A formal sum of these points
\[
D=P_1+P_2+P_3\,,\qquad \mbox{deg}D=3
\]
is a semi-reduced divisor of degree three on the elliptic curve  $E$ (\ref{e-curve}).

\subsection{Semi-reduced divisor of  degree two}
Following Abel's idea \cite{ab} let us consider variable points of intersection of
 elliptic curve $E$ with a family of curves depending on time
\[\Upsilon:\qquad \mathrm g(\mu, y) = c_2\mathcal Q(\mu)y-\mathcal P(\mu)=0\,.\]
Substituting
\[
y=\frac{\mathcal P(\mu)}{c_2\mathcal Q(\mu)}
\]
into the elliptic curve  equation (\ref{e-curve})
\[
f(\mu,y)=c_2^2y^2 + (c_1\mu^2 - H\mu + K)y + \det(A-\mu)=0
\]
we obtain Abel's polynomial
\[
\Psi=\theta\cdot B_3(\mu)\cdot B_2'(\mu)
\]
which determines an intersection divisor
\begin{equation}\label{dp-cl}
D+D'+D_\infty=(P_1+P_2+P_3)+(P_4+P_5)+D_\infty=0\,,
\end{equation}
where $D_\infty$ is a linear combination of the points at infinity.

Abscissas of points $P_4$ and $P_5$ are the roots of polynomial
\[\begin{array}{l}
B'_2(\mu)=b_2'(\mu-\mu_4)(\mu-\mu_5)=\Bigl(c_2^3-c_1c_2H  + (a_1 + a_2 + a_3)c_1^2c_2 - c_1^2X\Bigr)\mu^2 \\
\\
+\Bigl(c_1c_2K - (a_1a_2 + a_1a_3 + a_2a_3)c_2c_1^2 + (c_1H - 2c_2^2)X\Bigr)\mu + a_1a_2a_3c_1^2c_2 - c_1KX + c_2X^2\,.
\end{array}
\]
Ordinates of these points are equal to
\[
y_i=\frac{\mathcal P(\mu_i)}{c_2\mathcal Q(\mu_i)}\,,\qquad i=4,5.
\]
Affine coordinates of points $P_4$ and $P_5$  are functions on integrals of motion $c_1,c_2,H,K$ and  one variable $X$ (\ref{x-coord})   so that
\[
\{\mu_4,H\}\,\{\mu_5,K\}-\{\mu_4,K\}\,\{\mu_5,H\}=0\,,\qquad \{\mu_4,\mu_5\}\neq 0\,,
\]
and
\begin{equation}\label{line-2}
\mu_4-\mu_5=c_1(y_4-y_5)\,,
\end{equation}
i.e. points $P_4$ and $P_5$ move along a straight line and slope of this line is equal to $c_1^{-1}$.

\subsection{Reduced divisor of  degree one}
To directly apply the Euler \cite{eul0} and  Abel \cite{ab}  formulae we have to rewrite equation (\ref{e-curve}) in the following form
\[
z^2=\mathrm a_4\mu^4+\mathrm a_3\mu^3+\mathrm a_2\mu^2+\mathrm a_1\mu+\mathrm a_0
\]
using birational transformation
\[
y =z - \frac{c_1\mu^2 - \mu H+K}{2c_2^2}\,.
\]
Then  we have to consider various points of intersection of $E$ with a family of curves depending on time
\begin{equation}\label{par-cl}
\Upsilon':\qquad  z =\sqrt{\mathrm a_4}\mu^2+\mathrm b_1\mu+\mathrm b_0\,, \quad\mbox{where}\quad \mathrm a_4=\frac{c_1^2}{4c_2^4}\,.
\end{equation}
Here $\mathrm b_1$ and $\mathrm b_0$ are coefficients of the interpolating polynomial which is defined by equations
\[
z_4=\sqrt{\mathrm a_4}\mu_4^2+\mathrm b_1\mu_5+\mathrm b_0\,,\qquad
z_5=\sqrt{\mathrm a_4}\mu_5^2+\mathrm b_1\mu_5+\mathrm b_0\,,
\]
where $(\mu_4,z_4)$ and $(\mu_5,z_5)$  are abscissas and ordinates of  points $P_4$ and $P_5$ lying on  auxiliary curve $\Upsilon'$.
 The  corresponding  intersection divisor of curves $E$ and $\Upsilon'$ has the following form
\[
D'+D''+D_\infty=(P_4+P_5)+P_6+D_\infty=0\,.
\]
According to \cite{ab} abscissa $\mu_6$ of  point $P_6$ is equal to
\begin{equation}\label{mu6}
\mu_6=-\mu_4-\mu_5-\frac{2\mathrm b_0\sqrt{\mathrm a_4} + \mathrm b_1^2 - \mathrm a_2}{2\mathrm b_1\sqrt{\mathrm a_4} - \mathrm a_3}\equiv \frac{\nu}{\upsilon}\,.
\end{equation}
In the Clebsch case, it is a function on the integrals of motion with a numerator
\[\begin{array}{rcl}
\nu&=&c_1^6a_1a_2a_3 - c_1^5(a_1 + a_2 + a_3) K+ c_1^4\bigl(c_2^2(a_1 + a_2 + a_3)^2 + H\,K)\\
 \\
 &-&c_1^3\bigl(2(a_1+a_2+a_3)H - K\bigr)c_2^2 + c_1^2c_2^2\bigl(2c_2^2(a_1 + a_2 + a_3) + H^2) - 2c_1c_2^4H + c_2^6
\end{array}
\]
and denominator
\[
\begin{array}{rcl}
\upsilon&=&c_1^2\Bigl(c_1^4(a_1a_2 + a_1a_3 + a_2a_3)- c_1^3\bigl((a_1+a_2+a_3)H+K\bigr) + c_1^2\bigl(2(a_1 + a_2 + a_3)c_2^2 + H^2\bigr) \Bigr.
\\ \\
&-&\Bigl. 3c_1c_2^2H + 2c_2^4\Bigr)\,.
\end{array}
\]
Such reduced divisors also appear  for various superintegrable systems \cite{ts20rr, ts20m,ts20ds} and for the Kowalevski top \cite{ts20}.

Points $P_4, P_5$ and $P_6$ lie on   parabola $\Upsilon'$ of a constant size, which leads to the vanishing of determinant
\[
\left|
\begin{array}{ccc}
\mu_4 & z_4 & 1 \\
  \mu_5 & z_5 & 1 \\
  \mu_6 & z_6 & 1
\end{array}
\right|=0
\]
and to the vanishing of the corresponding Abel's integral relation \cite{ab}
\[
\frac{d\mu_4}{z_4}+\frac{d\mu_5}{z_5}+\frac{d\mu_6}{z_6}=0\,.
\]
Because abscissa $\mu_6$ (\ref{mu6}) is a constant of motion, i.e. $d\mu_6=0$, we immediately obtain the following equation for  abscissas of points from the support of  semi-reduced divisor $D'=P_4+P_5$ (\ref{dp-cl})
\begin{equation}\label{dx-eq1}
\frac{d\mu_4}{z_4}+\frac{d\mu_5}{z_5}=0\,.
\end{equation}
Because points $P_1,P_2,P_3$ and $P_4,P_5$ belong to the intersection divisor (\ref{dp-cl}) we also have  equation
\[
\frac{d\mu_1}{z_1}+\frac{d\mu_2}{z_2}+\frac{d\mu_3}{z_3}=0
\]
involving a regular differential on the elliptic curve $E$. This equation is a consequence of the fact that the unique reduced divisor in this class of equivalent divisors is a constant of motion.

Variables $\mu_{4,5}$  depend on integrals of motion $c_1,c_2,H,K$ and the one time-dependent variable $X$ (\ref{x-coord}) and, therefore,
they satisfy geometric equation (\ref{line-2}).

Let us introduce  elliptic coordinates $u_{1,2}$ on $e^*(3)$  using  standard definition
\[
\frac{p_1^2}{z-a_1}+\frac{p_2^2}{z-a_2}+\frac{p_3^2}{z-a_3}=\frac{c_1(z-u_1)(z-u_2)}{(z-a_1)(z-a_2)(z-a_3)}=0\,,
\]
and the corresponding conjugated momenta, see  \cite{ts06}.  These coordinates satisfy the Abel's  equations
\begin{equation}\label{nab}
\begin{array}{rcl}
\frac{\dot{u_1}}{\sqrt{(u_1-a_1)(u_1-a_2)(u_1-a_3)(u_1^2-Hu_1+\tilde{K})}}+
\frac{\dot{u_2}}{\sqrt{(u_2-a_1)(u_2-a_2)(u_2-a_3)(c_1u_2^2-Hu_2+\tilde{K})}}&=&0\,,\\
\\
\frac{u_1\dot{u_1}}{\sqrt{(u_1-a_1)(u_1-a_2)(u_1-a_3)(c_1u_1^2-Hu_1+\tilde{K})}}+
\frac{u_2\dot{u_2}}{\sqrt{(u_2-a_1)(u_2-a_2)(u_2-a_3)(c_1u_2^2-Hu_2+\tilde{K})}}&=&4\,,
\end{array}
\end{equation}
where $\tilde{K}=K-2c_2X(t)$ is a function on time. If we know $X(t)$, we can try to solve these equations for $u_1$ and $u_2$.

This variable $X$ also appears in the Kobb \cite{kobb} and Harlamova \cite{har59} calculations. We will discuss this topic in the forthcoming publication.

\section{Neumann's problem}
\setcounter{equation}{0}
At the special case
\[c_1=p_1^2+p_2^2+p_3^2=1\,,\qquad c_2=p_1M_1+p_2M_2+p_3M_3=0\]
 the Clebsch system is equivalent to the well-studied Neumann's problem  \cite{neum59} describing the motion of a particle on the unit sphere in the field of quadratic potential, see algebro-geometric description of this system in textbooks \cite{dub85,mum84}.

Sphero-conical coordinates $u_1$ and $u_2$ on the sphere are defined through  equation
\[
\frac{p_1^2}{z-a_1}+\frac{p_2^2}{z-a_2}+\frac{p_3^2}{z-a_3}=\frac{(z-u_1)(z-u_2)}{(z-a_1)(z-a_2)(z-a_3)}=0\,,
\]
which implies that $c_1=\sum p_1^2=1$. Similar to elliptic coordinates in $\mathbb R^3$, these coordinates in  $\mathbb S^2$ are also orthogonal and only locally defined. They take values in the intervals
\[
a_1<u_1<a_2<u_2<a_3\,,
\]
so that
 \[
p_i=\sqrt{\dfrac{(u_1-a_i)(u_2-a_i)}{(a_j-a_i)(a_m-a_i)}}\,
,\qquad i\neq j\neq m\,.
\]
If $\pi_{1,2}$ are canonically conjugated momenta
\[\{u_1,\pi_1\}=\{u_2,\pi_2\}=1\]
with respect to the Poisson brackets (\ref{e3}), then
\[
M_i=\dfrac{2\varepsilon_{ijm}p_jp_m(a_j-a_m)}{\mu_1-\mu_2}\Bigl((a_i-\mu_1)\pi_1-(a_i-\mu_2)\pi_2\Bigr)\,.
\]
At $c_1=1$ and $c_2=0$ Hamiltonians (\ref{ham}) are the second order  polynomials in momenta
\[
H_0=\frac{4\varphi_1\pi_1^2}{\mu_1 - \mu_2} - \frac{4\varphi_2\pi_2^2}{\mu_1 - \mu_2} + \mu_1 + \mu_2\,,\qquad
K_0=\frac{4\mu_2\varphi_1\pi_1^2}{\mu_1 - \mu_2} - \frac{4\mu_1\varphi_2\pi_2^2}{\mu_1 - \mu_2} + \mu_1 \mu_2\,,
\]
where we for brevity denote
\[
\varphi_k=\det(A-\mu_k)=(a_3 - \mu_k)(a_2 - \mu_k)(a_1 - \mu_k)\,, \qquad k=1,2.
\]
In this partial case, spectral curve $\Gamma$ (\ref{s-curve}) is the genus two hyperelliptic curve which can be rewritten in the following form
\begin{equation}\label{s-curve0}
\Gamma_0:\qquad   \Bigl( \det(\mu-A)\,\chi\Bigr)^2=(\mu^2 - H_0\mu + K_0)\,\det(\mu-A)\,,\qquad \chi=-\lambda^{-1}\,.
\end{equation}
Below we will study the evolution of semi-reduced and reduced divisors on this curve.

At  $c_2=0$ cubic polynomial $A_3(\lambda)$ (\ref{a-gen}) becomes quadratic polynomial on $\chi=\lambda^{-1}$
\[
A_3(\chi)=\sqrt{\varphi_1\varphi_2}\,(\chi-2\pi_1)(\chi-2\pi_2).
\]
Two roots of this polynomial define  reduced divisor of degree two
\[D=P_1+P_2\,,\quad |D|=0,\qquad  P_1=(u_1,2\pi_1)\,,\quad P_2=(u_2,2\pi_2)\]
on the genus two spectral curve $\Gamma_0$ (\ref{s-curve0}).

 In this case we have not third degree semi reduced  divisor $D$ and, therefore, we can not apply Abel's reduction in this case .

 \subsection{Semi-reduced divisor}
At $c_2=0$ cubic polynomial $B_3(\mu)$ (\ref{b-gen}) remains a cubic polynomial on $\mu$
\begin{equation}\label{U0}
B_3(\mu)=b_3(\mu-\mu_1)(\mu-\mu_2)(\mu-\mu_3)=\frac{4(\mu - u_1)(\mu - u_2)^2\varphi_1\pi_1^2}{u_1 - u_2}
 -\frac{4(\mu - u_1)^2(\mu-u_2)\varphi_2\pi_2^2}{u_1 - u_2}\,.
\end{equation}
Three roots of this polynomial $\mu_1,\mu_2$ and $\mu_3$ define a semi-reduced divisor of poles  on the genus two spectral curve $\Gamma_0$ (\ref{s-curve0})
\[
D=P_1+P_2+P_3\,,\qquad P_1=(u_1,2\pi_1)\,,\qquad P_2=(u_2,2\pi_2)\,,\qquad P_3=(\mu_3,\chi_3)\,,
\]
where affine coordinates of the  third pole are functions on  variables $u_{1,2}$ and $,\pi_{1,2}$
\begin{equation}\label{u3-0}
\mu_3\equiv\frac{K_0-u_1u_2}{H_0 - u_1 - u_2}\,,\qquad \qquad
\chi_3=\frac{4\pi_1\pi_2}{H_0-u_1-u_2}\sqrt{\frac{\varphi_1\varphi_2}{\varphi_3}}
\end{equation}
Evolution of three poles $P_1,P_2$ and $P_3$ along the curve $\Gamma_0$ (\ref{s-curve0}) is determined by Abel's equations
\begin{equation}\label{eqm-0}
\Omega_1(P_1)\dot{\mu}_1+\Omega_1(P_2)\dot{\mu}_2=0\qquad
\Omega_2(P_1)\dot{\mu}_1+\Omega_2(P_2)\dot{\mu}_2=4\,,\quad
\end{equation}
and
\[
\Omega_1(P_3)\dot{\mu}_3=\frac{4X}{\sqrt{\varphi_3}(H_0 - u_1 - u_2)}\,,
\]
where regular differentials on  $\Gamma_0$ (\ref{s-curve0}) have the form
\[
\Omega_1=\frac{1}{\det(\mu-A)\,\chi}\,,\qquad
\Omega_2=\frac{\mu}{\det(\mu-A)\,\chi}\,,
\]
and variable $X$  (\ref{x-coord}) at $c_2=0$ is equal to
\[
X=a_1p_1M_1 +a_2p_2 M_2 +a_2p_3M_3=\frac{2\sqrt{\varphi_1\varphi_2}(\pi_1-\pi_2)}{u_1-u_2}\,.
\]
 Abscissa and ordinate of the third point $P_3$ are functions of elliptic coordinates and integrals of motion, and, therefore,  two Abel's equations (\ref{eqm-0}) completely determine the evolution of the semi-reduced divisor of poles
\begin{equation}\label{d-neu}
D=P_1+P_2+P_3\,,\qquad \mbox{deg}D=3\,,\qquad \mbox{dim}|D|= \mbox{deg}D-g=3-2=1\,.
\end{equation}
\subsection{Reduced divisor}
According to the Riemann-Roch theorem, there is a unique reduced divisor $D'$ on  genus two hyperelliptic curve $\Gamma_0$ (\ref{s-curve0})
\[
\rho(D)=D'=P_4+P_5\,,\qquad \mbox{deg}D'=2\,,\qquad \mbox{dim}|D'|= \mbox{deg}D'-g=2-2=0\,,
\]
which is equivalent to semi-reduced divisor $D$ (\ref{d-neu}).

Following  Abel \cite{ab}, we can identify reduced divisor  $D'$ with a part of the  intersection divisor
\[
D+D'+D_\infty=0
\]
of   hyperelliptic curve $\Gamma_0$  with  a family of parabolas $\Upsilon$ involving three points $P_1,P_2$  and $P_3$ on a projective plane. Parabola \[\Upsilon:\,y=V_0(\mu)\]
is defined by interpolation polynomial
\begin{equation}\label{V0}
V_0(\mu)=\frac{(\mu-\mu_2)(\mu-\mu_3)}{\mu_1-\mu_2)(\mu_1-\mu_3)}\chi_1+
\frac{(\mu-\mu_1)(\mu-\mu_3)}{\mu_2-\mu_1)(\mu_2-\mu_3)}\chi_2+
\frac{(\mu-\mu_1)(\mu-\mu_2)}{\mu_3-\mu_1)(\mu_3-\mu_2)}\chi_3\,.
\end{equation}
Substituting $y=V_0(\mu)$ into  (\ref{s-curve0}) we obtain Abel's polynomial $\Psi$ defining abscissas $\mu_{4,5}$ of  points  $P_4$ and $P_5$
\[
\Psi=\theta\cdot B_0(\mu)\cdot B_0'(\mu)\,,\qquad B_0=b_3(\mu-\mu_1)(\mu-\mu_2)(\mu-\mu_3)\,,\quad B_0'=(\mu-\mu_4)(\mu-\mu_5)\,,
\]
see  Abel's original calculations for genus two hyperelliptic curves \cite{ab} or \cite{ts15} and references within. In our case
\[\begin{array}{l}
B_0'(\mu)=\mu^2+(\mu_1 + \mu_2 + \mu_3 - a_1 - a_2 - a_3 - H_0)\mu+\mu_1^2 + \mu_2^2 + \mu_3^2\\ \\
 - (a_1 + a_2 + a_3)(\mu_1 + \mu_2 + \mu_3)+(a_1 + a_2 + a_3 - \mu_1 - \mu_2)H_0 + a_1a_2 + a_1a_3 + a_2a_3 + 2\mu_1\mu_2\,,
\end{array}
\]
where
\[
\mu_1=u_1\,,\qquad  \mu_2=u_2\,,\qquad \mu_3=\frac{K_0-u_1u_2}{H_0 - u_1 - u_2}\,.
\]
Affine coordinates of this reduced divisor $D'=P_4+P_5$  satisfy the same Abel's equations (\ref{eqm-0})
\[
\Omega_1(P_4)\dot{u}_4+\Omega_1(P_5)\dot{u}_5=0\quad\mbox{and}\quad
\Omega_2(P_4)\dot{u}_5+\Omega_2(P_5)\dot{u}_5=4\,,\quad
\]
but
 \[
\{\mu_4,\mu_5\}\neq 0\,.
\]
In this case affine coordinates of reduced divisor have  properties similar to properties of  reduced divisor on genus two algebraic curve for Euler's two fixed centers problem  \cite{ts21}.

\section{Conclusion}
We study properties of the  $n\times n$ Lax matrix obtained by Perelomov \cite{p81}.  When $n=3$ Lax equations coincide with equations of motion for the Clebsch system. We study reduction of divisors  on the corresponding elliptic spectral curve
\[D= P_1+P_2+P_3\to D'=P'_1+P'_2\to \rho(D)=P''\]
using standard Abel's reduction algorithm \cite{ab}.   We prove that reduced divisor is fixed, i.e. independent on time, when dimension of the configurational space is more then genus of the corresponding spectral curve.

Affine coordinates of semi reduced divisor $D'$ are functions on integrals of motion and one dynamical variable $X$, which is related to quadratures by Kobb \cite{kobb} and Harlamova \cite{har59}. Evolution of this semi reduced divisor and its relations with  known quadratures will be discussed later.

\section*{FUNDING}

This work of  A.V. Tsiganov was supported by the Russian Science Foundation (project no.~19\--\-71\--\-30012) and performed at the Steklov Mathematical Institute of the Russian Academy of Sciences.

\section*{CONFLICT OF INTEREST}
The authors declare that they have no conflicts of interest.

\end{document}